\documentclass[11pt,onecolumn]{article}
\usepackage[utf8]{inputenc}

\usepackage{mathtools}
\usepackage{amsmath}
\usepackage{amssymb}
\usepackage{amsthm}
\usepackage{amsfonts}
\usepackage{geometry}
\usepackage{graphicx}
\usepackage[dvips]{epsfig}
\usepackage{braket}
\usepackage[affil-it]{authblk} 
\usepackage{hyperref}
\usepackage{cite}
\usepackage{placeins}

\begin{document}
\title{Noncommutative Deformation of Optical States}

\author{Aamir Rashid\thanks{\href{mailto:aamirjamian@pondiuni.ac.in}{aamirjamian@pondiuni.ac.in}}}
\author{Jishnu Aryampilly\thanks{\href{mailto:jishnu.sankaran@pondiuni.ac.in}{jishnu.sankaran@pondiuni.ac.in}}}

 \affil{Department of Physics, \\Pondicherry University, \\ Puducherry 605014, India}

\date{Dated: \today}

\maketitle

\section{Abstract}
In this paper, we  study the noncommutative deformation of different optical states. We develop the deformed coherent state by using the raising and lowering operators of the quantum harmonic oscillator. This helps us to investigate the noncommutative deformation of a squeezed state in terms of noncommutative parameter, which in turn leads to the noncommutative deformation of a photon-added coherent state. This noncommutative deformation have an overall effect on the non-classicality of such states. This is done by investigating the effect of noncommutative deformation of the Mandal parameter for such states.

PACS number(s): 11.10.Nx, 03.65.-w

\section{Introduction}
In the search of the unification of gravity and quantum mechanics, noncommutative geometry is playing a substantial role \cite{Doplicher:1994tu, Douglas:2001ba}. It is expected that the quantization of space-time will modify its short-distance behavior and such a
quantum space-time would be described by noncommutative geometry\cite{Witten:1985cc}. The concept of noncommutativity was given in various contexts like gravity theories \cite{Aschieri:2005yw, Calmet:2005qm, Harikumar:2006xf, Balachandran:2006qg} and quantum mechanics \cite{Muthukumar:2006ab, Muthukumar:2007zza, Muthukumar:2004wj}. Recently the noncommutative formulation of quantum mechanics has been put forward in which the main ingredient is the construction of coherent states in Hilbert space of Quantum Mechanics \cite{BenGeloun:2009zd}. Noncommutative background manifolds have been considered in field theories. This has motivated the study of 
noncommutative deformation of various quantum systems and such a deformation has produced interesting modifications to such quantum systems. Some systems have originated from noncommutative quantum mechanical structures, and in those systems, the development of coherent states has been studied \cite{Dey:2018bmr} . The photon-added coherent states of a noncommutative harmonic oscillator associated with a deformed oscillator algebra
have been studied in \cite{Chowdhury:2012ik}. Various coherent states have been studied in noncommutative spaces \cite{Dey:2012tv, Dey:2012pq, Ghosh:2011ze}, in that study, the nonclassical states \cite{Dey:2018bmr, Glauber:1962tt, Sudarshan:1963ts}, which originate from coherent states \cite{Dey:2018bmr} of the system are also discussed. Further more other states like photon added coherent states \cite{Dey:2016vcm,Sivakumar:1998yn} and their squeezing and squeezed states  have been studied. This included the effect of noncommutativity on the Mandel parameter. The self-similarity properties of fractals have been investigated using deformed algebra of coherent states.

The noncommutative deformation of various systems has been studied using Moyal star product \cite{Connes:1994yd, Nair:2000ii, Nicolini:2004yb, Smailagic:2003rp}. The coherent states over noncommutative spaces have been discussed in \cite{Jing:2005db, Yin:2005ij} through deformed Heisenberg-Weyl algebras. The effect of non-commutativity on the minimal uncertainty states of a deformed quantum system has been thoroughly studied. It has been argued that photon added coherent states and their dynamics are modified by noncommutative deformation of the system \cite{28,29,30,31}. It has also been observed that such photon-added coherent states have interesting properties  \cite{32,33,34}. Non-linear properties of photon added coherent states have also been studied  \cite{35,36}.  Thus, it would be interesting to analyze the noncommutative deformation of quantum optical states. In this paper, we will analyze the deformation of quantum optical states.
\section{Coherent States}
In this section, we will develop the noncommutative deformation of coherent states. Discussion on coherent states over noncommutative spaces also appears in literature through deformed Heisenberg-Weyl algebras \cite{37} .
 Thus, we can propose that , Heisenberg algebra is deformed as $
[x_1,x_2]=i\theta 
$, 
where $x_1$ and $x_2$ are the usual  spatial coordinates. 
Now we can define $a$ and $a^\dagger$ as   
\begin{equation}
a=x_{1}+ix_{2},\quad  a^{\dagger}=x_{1}-ix_{2}
\end{equation}
such that the commutation between them is $
[a,a^{\dagger}]=2\theta
$. 

The coherent state \cite {x1} for these operators can now be defined through displacement operator, expressed as 
\begin{equation}
\ket\alpha=D(\alpha)\ket 0,
\end{equation}
where  $ D(\alpha)$ which is a unitary operator, and can be obtained from $a$ and $a^\dagger$ as  
\begin{equation}
D(\alpha)=e^{\frac{1}{\theta}(\alpha^{*}a-\alpha a^{\dagger})}.
\end{equation}
Here $\alpha $ is a complex number.
Now using the noncommutative  deformation of this system with  $ \theta$ taken as the  parameter of noncommutativity, and using the  Baker-Campbell-Hausdroff formula,  we can write 
\begin{eqnarray}
\ket\alpha &=& e^{\frac{\alpha a}{\theta}} e^{\frac{-\alpha a^\dagger}{\theta}}e^{\frac{-\alpha^{2}2\theta}{2\theta^{2}}}\ket 0,
\nonumber \\ &=&  e^{\frac{-\alpha^{2}}{\theta}}e^{\frac{-\alpha a^\dagger}{\theta}}\sum_{n=0}^{n=\infty}
{\frac{ \alpha^n a^n}{\theta^{n}n!}}\ket 0,
\nonumber \\ &=& e^{\frac{-\alpha^2}{\theta}}\sum_{n=0}^{n=\infty}{\frac{(-1)^n\alpha^n}{\theta^{n}\sqrt(n!)}}\ket n.
\end{eqnarray}
This defines a coherent state in noncommutative  space.  

A  useful parameter of non-classicality  is the  Mandel parameter \cite{38}.  We can calculate the Mandel parameter of this system as
\begin{align}
Q_M &={\frac{<(a^{\dagger}a)^{2}>-<{a^{\dagger}a>}^{2}-<{a^{\dagger}a>}}{<{a^{\dagger}a>}}}\nonumber \\
&={\frac{ 2\theta\alpha^{2}+\alpha^{4}-<\alpha^{2}>^{2}-{\alpha^{2}}}{\alpha^{2}}}\nonumber\\
&={2\theta-1} 
\end{align}
Thus, we observe the noncommutative  deformation has an effect on the  non-classicality  of coherent states. 

We also observe that the probability of detecting photons in $n$ state can be written as 
\begin{align}
P_n &=|<n|\alpha>|^{2}\nonumber\\
&=|<n|n>|^{2}e^{\frac{(-\alpha)^{2}}{\theta^{2}}}{\frac{(-1)^{2n}(\alpha)
^{2n}}{(\theta)^{2n}n!}}\nonumber\\
&=e^{\frac{(-\alpha)^{2}}{\theta^{2}}}{\frac{(\alpha)^{2n}}{(\theta)^{2n}n!}}
\end{align}
Now we observe that the photon distribution is a Poisson distribution, even after noncommutative effects have been considered.  This usual deformation is deformed by the  $\ theta $ parameter, as is expected. However, we have obtained the explicit form of such a deformed distribution.  
\subsection{Single Mode Squeezed States}

The concept of squeezed states is related to coherent states developed from the examples of oscillations of the electromagnetic field in the pioneering works by Glauber, Klauder, and Sudarshan in \cite{Bargmann:1971ay}. These states are constructed using a suitable unitary operator.Deformed squeezed states are introduced in \cite{Aneva:2002jna}. 
Now we can analyze the effects of noncommutative deformation on single-mode squeezed states and two-mode squeezed states. The simplest squeezed states are obtained from the vacuum state by applying the squeezing operator, which can be expressed as 
\begin{equation}
S(\zeta)=e^{\frac{\zeta a\dagger^{2}}{2}-\frac{\zeta a^{2}}{2}}
\end{equation}
Here $\ket\zeta$ is defined as 
\begin{equation}
\ket\zeta=e^{\frac{\zeta a\dagger^{2}}{2}}e^{-\frac{\zeta a^{2}}{2}}e^{[{\frac{\zeta a\dagger^{2}}{2},{-\frac{\zeta a^{2}}{2}}}] }\ket0
\end{equation}

Now using the  Baker-Campbell-Hausdroff formula, we can write   
\begin{align}
\ket\zeta &=e^{\frac{\zeta a\dagger^{2}}{2}}e^{\frac{-\zeta a^{2}}{2}} e^{{\frac{\zeta^{2}}{2} {(a,a\dagger)(a\dagger a+aa\dagger)}} }\ket 0\nonumber \\
&=e^{\frac{\zeta a\dagger^{2}}{2}}e^{-\frac{\zeta a^{2}}{2}}e^{\zeta^{2}\theta[a\dagger a+2\theta+a\dagger a] }\ket 0
\end{align}
Now we can express this as 

\begin{align}
\ket\zeta &=e^{\frac{\zeta {a^{\dagger}}^{2}}{2}}e^{\frac{-\zeta a^{2}}{2}}[1+2\zeta^{2}\theta(a\dagger a+\theta)] \ket  0
\nonumber \\
&=e^{\frac{\zeta {a^{\dagger}}^{2}}{2}}e^{\frac{-\zeta a^{2}}{2}}[\ket 0+2\zeta^{2}\theta^{2}\ket 0]
\end{align}
Finally, we can write this state as 
\begin{align}
\ket\zeta &= e^{\frac{\zeta {a^{\dagger}}^{2}}{2}}[\ket 0+2\zeta^{2}\theta^{2}\ket 0]\nonumber \\
&= [1+2\zeta^{2}\theta^{2}]e^{\frac{(\zeta) {a^{\dagger}}^{2}}{2} }\ket 0 \nonumber \\
&= [1+2\zeta^{2}\theta^{2}]\sum_{n=0}^{n=\infty}{\frac{\zeta^{n} ({a^{\dagger}}^{2})^{n}}{2^{n}n!}}\ket 0 \nonumber\\
&= [1+C\zeta^{2}\theta^{2}]\sum_{n-0}^{n=\infty}{\frac{\zeta^{n}{a^{\dagger}}^{n}}{2^{n}\sqrt{n}!}}\ket n 
\end{align}

where $C$ is a constant. 
 The Fock state representation is obtained as 
\begin{equation}
\ket\zeta=2\theta\sqrt{sechr}\sum\frac{-1}{2}{e^{i\phi}} {tanhr}^n\frac{\sqrt{(2n)!}}{n!}\ket{ 2n}
\end{equation}
Thus, we have explicitly obtained the noncommutative deformation of the squeezed state. Here we again observe that the deformation is proportional to the noncommutative parameter $\theta$.

\subsection{Two Mode Squeezed States}
Denoting the two  modes by a and b respectively, we can write their commutators as
\begin{equation}
[a,a^{\dagger}]=2\theta,\quad [b,b^{\dagger}]=2\theta,
\end{equation}
 where , 
 \begin{equation}
 b=x_3+ix_4, \quad b^{\dagger}=x_3-ix_4 .
\end{equation}
 
Taking $ \theta_{12}=\theta_{34}=\theta$,    $ \theta_{21}=\theta_{43}=-\theta $.
we can write the  squeezed states by applying the two mode squeezing operator  $ S(\zeta) $ on  two mode vaccum $ |0, 0\rangle $ where $ S(\zeta) = e^{{\zeta a\dagger b^\dagger}-{\zeta^{*} ab}} $, So that two mode squeezed state is expressed as
\begin{equation}
\ket\zeta=e^{{\zeta a\dagger b^\dagger}-{\zeta^{*} ab}} |0,0\rangle ,
\end{equation}
\begin{align}
\ket\zeta &= e^{\zeta a\dagger b^\dagger}[1+2\zeta^{2}\theta^{2}][|0,0\rangle],
\end{align}
\begin{align}
\ket\zeta=[1+C\zeta^{2}\theta^{2}]\sum_{n=0}^{n=\infty}{\frac{\zeta^{n}{a^{\dagger}}^{n}{b^\dagger}^n}{{n}!2^n}} [|0,0\rangle].
\end{align}
The Fock state representation is given by the expression 
\begin{equation}
 \ket\zeta= 2\theta{sechr}\sum_{n=0}^{n=\infty}{e^{in\phi}} {tanhr}^n\ket{ n,n}.
\end{equation}

Thus we have derived the noncommutative deformation of two-mode squeezed state.
\subsection{Photon Added Coherent States}
A photon-added coherent state is obtained from one-photon excitations of a classical coherent state. Such states can be viewed as states between coherent states and Fock states.  We will analyze the noncommutativity deformation of such states.  The photon added coherent states $ \ket {\alpha_0, m} $ are defined as 
\begin{equation}
 \ket {\alpha_0,m}=\frac{a^{\dagger m}\ket{\alpha_0}}{\sqrt{<\alpha_0|a^{m}a^{\dagger m}|\alpha_0>}}.
\end{equation}
Now for such  photon added coherent states, we can express  expectation values  $<a>_{c}^{a}$ as  
\begin{eqnarray}
<a>_{c}^{a}&=&\frac{<\alpha_0|aaa^\dagger| \alpha_0>}{1+\alpha^{2}},
\nonumber \\
 &=&\frac{6 \theta\alpha_0^{2}+\alpha_0^{4}}{1+\alpha_0^{2}}.
\end{eqnarray}
We can also express $<a^\dagger a>_c^a$ as 
\begin{align} 
<a^\dagger a>_c^a&=\frac{2\theta[1+\alpha_0^{2}]}{1+\alpha_0^{2}},\\ 
 &= \frac{<\alpha_0|aa^{\dagger{2}} a^ {2} a\dagger|\alpha_0>}{1+\alpha_0^{2}} \nonumber .
\end{align}
So, we can write the Mandel parameter  for such states as 
\begin{align}
Q_M&={\frac{<(a^{\dagger}a)^{2}>-<{a^{\dagger}a>}^{2}-<{a^{\dagger}a>}}{<{a^{\dagger}a>}}},\nonumber \\
&=\alpha_0^{2}-\frac{2\theta}{1+\alpha_0^{2}}-1\nonumber, \\
&=\frac{\alpha_0^{4}-2\theta-1}{1+\alpha_0^{2}}.
\end{align}
It may be noted that the Mandal parameter for these states is again deformed by the noncommutative parameter   $\theta$. 
The noncommutative deformation has an effect on the  non-classicality of photon added coherent states. 
Hence, we obtained a noncommutative deformation of photon added coherent states. 
\subsection{Conclusion}
In this paper we analyzed the noncommutative deformation of different optical states. This was done by deforming the commutative Heisenberg algebra. This in turn modified the relation between the creation and annihilation operators. These deformed creation and annihilation operators were first used to define a deformed coherent state. Then they were used to deform a squeezed state. The additional factor of noncommutative structure can point towards the noncommutative structure of spacetime at a high energy regime in the opto-quantum mechanical experimental setup. This can lead to a route towards the realization of the phenomena of quantum gravity in a quantum optical experiment.
 Finally, such deformation of a photon added coherent states was derived. We analyzed the effect of noncommutative deformation on the non-classicality of such states. This was done by 
analyzing the effect of noncommutative deformation on Mandal parameter for such states. 
It would be interesting to study such deformation of cat states. Then those deformed cat states can be used to investigate  the effect of noncommutative deformation on non-classicality  of deformed cat states. This can be done by analyzing the effect of non-commutativity on the Mandal parameter for these deformed cat states. It would also be interesting to analyze other deformation of such optical systems. These could include various other deformations of the Heisenberg algebra. 
\section{Acknowledgement}
This work was motivated  by the discussion with    Mir Faizal https://orcid.org/0000-0002-3292-9426,
Irving K. Barber School of Arts and Sciences, University of British Columbia — Okanagan, Kelowna, British Columbia, Canada
Canadian Quantum Research Center, Vernon, British Columbia, Canada .

\end{document}